\documentclass[11pt]{article}
\usepackage{moriond,epsfig}

\def\be{\begin{equation}}
\def\ee{\end{equation}}
\def\bea{\begin{eqnarray}}
\def\eea{\end{eqnarray}}

\def\lsim{\mathrel{\raise.3ex\hbox{$<$\kern-.75em\lower1ex\hbox{$\sim$}}}}
\def\simlt{\stackrel{<}{{}_\sim}} \def\simgt{\stackrel{>}{{}_\sim}}
\begin{document}
\vspace*{4cm}
\title{THE FINE-TUNING PROBLEM IN LITTLE HIGGS MODELS}

\author{I. Hidalgo}

\address{IFT-UAM/CSIC, Facultad de Ciencias, C-XVI \\
Universidad Aut\'onoma de Madrid, 28049 Madrid, Spain}

\maketitle\abstracts{
Little Higgs models represent an alternative to Supersymmetry as a solution to the Hierarchy Problem. After introducing the main physical ideas of these models, we present the fine-tuning associated to the electroweak breaking in Little Higgs scenarios. Taking into account the most general properties of this scenarios and focusing on two representative ``Little Higgs'' models, we find that the fine-tuning is much higher than suggested by the rough estimates usually made. The main sources that increase the fine-tuning in these models are identified, then they can be taken into account in order to construct a successful model.}

\section{Introduction}

The Standar Models (SM) describes accurately almost all particle physics experiments. Despite its remarkable success, it is commonly assumed that the Big Hierarchy problem of the SM motivates the existence of New Physics (NP) beyond the SM at the scale $\Lambda_{\rm SM}\simlt$ few TeV. The Big Hierarchy problem is based on the sensitivity of the Higgs mass to quadratic divergences \cite{veltman}. At one-loop these divergences are:
\be
\label{quadrdiv}
\delta_{\rm q} m^2 = {3\over 16\pi^2 v^2}(2m_W^2 + m_Z^2 + m_h^2 -
4 m_t^2)\Lambda_{\rm SM}^2\ , 
\ee
where $m_W^2=\frac{1}{4}g^2v^2$, $m_Z^2=\frac{1}{4}(g^2+g'^2)v^2$
and $m_t^2=\frac{1}{2}\lambda_t^2v^2$. 
Hence the requirement of no fine-tuning between the 
above contribution
and the tree-level value of $m^2$ sets an upper bound on
$\Lambda_{\rm SM}$. E.g. for $m_h = 115-200$ GeV
\be
\label{quadrft}
\left|{\delta_{\rm q} m^2 \over m^2}\right| \leq 10\ \Rightarrow \
\Lambda_{\rm SM}  \simlt 2-3\ {\rm TeV}   \ee

The previous upper bound on $\Lambda_{\rm SM}$ is in some tension with
the experimental lower bounds on the suppression scale of higher order
operators, which typically give $\Lambda_{\rm LH}\simgt$ 10 TeV.  This is known as the ``Little Hierarchy'' problem.

\section{Fine-tuning analysis in Little Higgs Models}

Little Higgs (LH) models \cite{LH} are a recent development built as a solution of the Little Hierarchy problem. These models stabilize the Higgs mass by making the Higgs a pseudo-Goldstone boson resulting from a spontaneously broken global symmetry. There is also a explicit breaking of this global symmetry done by Yukawa and gauge couplings in a collective way. Therefore, the SM Higgs mass is protected at one-loop from quadratically divergent contributions, what is in principle enough to avoid the Little Hierarchy problem: if quadratic corrections to $m^2$ appear at two-loops, then without fine-tuning price we can have a cut-off of 10 TeV.

In LH models the fine-tuning associated to electroweak breaking should be checked in practice and must be computed with the same level of rigor employed for the supersymmetric models. In this paper the analysis is done on two particular models: the Littlest Higgs and the Simplest Little Higgs. To quantify the fine-tuning we follow Barbieri
and Giudice \cite{BG}: we write the Higgs VEV as $v^2=v^2(p_1, p_2,
\cdots)$, where $p_i$ are initial parameters of the model under study,
and define $\Delta_{p_i}$, the fine tuning parameters associated to
$p_i$, by
\be
\label{ftBG}
{\delta M_Z^2\over M_Z^2}= {\delta v^2\over v^2} = \Delta_{p_i}{\delta
p_i\over p_i}\ ,  \ee
where $\delta M_Z^2$ (or $\delta v^2$) is the change induced in
$M_Z^2$ (or $v^2$) by a change $\delta p_i$ in $p_i$. Roughly speaking
$|\Delta^{-1}_{p_i}|$ measures the probability of a  cancellation
among terms of a given size to obtain a result which is
$|\Delta_{p_i}|$ times smaller. The total fine-tuning is given by
\be
\label{Deltatot}
\Delta \equiv\left[\sum_i\Delta_{p_i}^2\right]^{1/2}\ .  \ee

It is important to note that the Little Hierarchy problem is itself a fine-tuning problem \cite{FT1,FT2}. E.g. one
could simply assume $\Lambda_{\rm SM}\simgt 10$ TeV with the 'only'
price of tuning $\delta_{\rm q} m^2$, as given by
eq.(\ref{quadrdiv}), at the 0.4--1 \% level (or, equivalently,
$\Delta = 100-250$).

In this paper we are going to focus on two particular LH models: the Littlest Higgs and the Simplest Little Higgs.

\subsection{The Littlest Higgs}

The Littlest Higgs model \cite{Littlest} is a non-linear sigma model based on a global
$SU(5)$ symmetry which is spontaneously broken to $SO(5)$ at a scale
$f\sim 1$ TeV, and   explicitely broken by the gauging of an
$[SU(2)\times U(1)]^2$ subgroup.  After the spontaneous breaking, the
latter gets broken to its diagonal subgroup, identified with the SM
electroweak gauge group, $SU(2)_L\times U(1)_Y$.   From the 14
(pseudo)-Goldstone bosons of the $SU(5)\rightarrow SO(5)$ breaking, 4
degrees of freedom (d.o.f.) are true Goldstones and the remaining 10 d.o.f. correspond to the SM Higgs
doublet, $H=(h^0,h^+)$, (4 d.o.f.) and a complex $SU(2)_L$ scalar
triplet, $\phi$ (6 d.o.f.) with $Y=1$;  in vectorial notation,
$\phi=(\phi^{++},\phi^+,\phi^0)$. 

The $[SU(2)\times U(1)]^2$ gauge interactions give a radiative mass to
the SM Higgs, but only when the couplings of both groups are
simultaneously present. Hence, the
quadratically divergent contributions only appear at two-loop order. And this mechanism is also used for potentially dangerous top-Yukawa interactions. The relevant states besides those of the SM are: the
pseudo-Goldstone bosons $H,\phi$; the heavy gauge bosons, $W', B'$, of
the axial $SU(2)\times U(1)$; and the two extra fermionic d.o.f. that combine in a vector-like heavy ``Top", $T$. The Lagrangian has a kinetic part and a fermionic one:
\be \label{L1} 
{\cal L} = {\cal L}_{kin}(g_1, g_2, g_1', g_2')\ +\ {\cal L}_{f}(\lambda_1, 
\lambda_2)  \ , \ee 
where $g_1, g'_1$ ($g_2, g'_2$) are the gauge 
couplings of the first (second) $SU(2)\times U(1)$ factor, and $\lambda_1, 
\lambda_2$ are the two independent fermionic couplings. These couplings 
are constrained by the relations with the SM couplings when we identified the diagonal subgroup with the SM electroweak group.

This Lagrangian gives ${\cal O}(f)$ masses to $W', B'$ and $T$. At this level, $H$ and $\phi$ are massless, but they get massive
radiatively. We must consider the quadratically divergent
contribution  to the one-loop scalar potential, given by \be
\label{V1gen}
V_1^{\rm quad} = {1\over 32 \pi^2}\Lambda^2\ {\rm Str} {\cal M}^2 \ ,
\ee where the supertrace ${\rm Str}$ counts degrees of freedom with a
minus sign for fermions, and ${\cal M}^2$ is the (tree-level,
field-dependent) mass-squared matrix. In this case $V_1^{\rm quad}$ does not contain a mass term for $h$, but these symmetries do not protect the mass of the triplet. In fact, if we consider gauge and fermion loops one sees 
that the Lagrangian should also include gauge invariant terms of the form,
\be 
\label{DeltaL1} -\Delta{\cal L} = c\ {\cal 
O}_V(\Sigma) \ +\ c'\ {\cal O}_F(\Sigma)\ , \ee 
with $c$ and $c'$ assumed to be constants of ${\cal O}(1)$. These operators produce a mass term for the triplet $\phi$ of order $\Lambda^2/(16  \pi^2)\sim f^2$. For fine-tuning analysis we need to know the natural size
of  $c$ and $c'$. Computing the one-loop contributions to  $c$ and $c'$
coming from (\ref{V1gen}) we get
\bea
\label{ccp}
c &=& c_0 \ +\ c_1 = c_0 \ + 3/4\ , \nonumber\\ c' &=& c_0' \ +\ c_1'
= c_0' \ - {24}\ .  \eea where the subindex 0 labels the unknown
threshold contributions from  the physics beyond $\Lambda$.

Besides giving a mass to $\phi$, the operators in eq.~(\ref{DeltaL1})
produce a quartic coupling for $h$. After integrating out the triple, the Higgs
quartic coupling $\lambda$ can be written in  the simplest manner as
\be
\label{lambda}
{1\over \lambda}={1\over \lambda_a}+{1\over \lambda_b}\ , \ee
with $\lambda_a\equiv c(g_2^2+g_2'^2)-c'\lambda_1^2$ and
$\lambda_b\equiv c(g_1^2+g_1'^2)$. 

Finally, a non-vanishing mass parameter for $h$ arises from the
logarithmic and finite contributions to the effective potential. In
the $\overline{\rm MS}$ scheme, setting the renormalization scale
$Q=\Lambda$,
\bea
\label{m2}
m^2 &=& {3\over 64\pi^2}\left\{3g^2M_{W'}^2\left[\log{\Lambda^2\over
M_{W'}^2} + {1\over 3}\right] + g'^2M_{B'}^2\left[\log{\Lambda^2\over
M_{B'}^2}+ {1\over 3}\right] \right\} \nonumber\\ &+&{3\lambda\over
8\pi^2}M_\phi^2 \left[\log{\Lambda^2\over  M_\phi^2}+ 1\right]
-{3\lambda_t^2\over 8\pi^2}M_T^2 \left[\log{\Lambda^2\over M_{T}^2}+
1\right]\ , \eea
where we have included the contribution from the $\phi$ masses. Therefore, the effective potential of the Higgs field can be written
in  the SM-like, with $ v^2=-{m^2\over \lambda}$  where $\lambda$ and $m^2$ are given by eqs.~(\ref{lambda}) and
(\ref{m2}).

A rough estimate of the fine-tuning associated to electroweak breaking
in the Littlest Higgs model can be obtained from  eq.~(\ref{m2}). The
contribution of the heavy top, $T$, to the Higgs mass is $\delta_T m^2 \geq 0.37 f^2$. Thus the
ratio $\delta_T m^2 / m^2$, tends to be quite large: {\it e.g.} for
$f= 1$ TeV and $m_h=115, 150, 250$ GeV, $\delta_T m^2 / m^2\geq 56,\
33,\ 12$. Since  there are other potential sources of fine-tuning,
this should be considered as a lower bound on the total
fine-tuning.

In order to perform a complete fine-tuning analysis \cite{FT2} we determine first
the independent parameters, $p_i$, and then calculate the associated
fine-tuning parameters, $\Delta_{p_i}$. For the
Littlest Higgs model the input parameters are $g_1$, $g_2$, $g_1'$, $g_2'$, 
$\lambda_1$, $\lambda_2$, $c_0$,$c'_0$ and $f$. We have not included $\Lambda$ because we are
assuming $\Lambda \simeq 4\pi f$. We can also ignored $\Delta_f$ because the parameter $f$ basically appears as a multiplicative factor
in the mass parameter, $m^2$, so $\Delta_f$ is always ${\cal O}(1)$. With this information we can calculate the total fine-tuning in the Littlest Higgs. This fine-tuning depends strongly on the region of parameter space considered and decreases significantly as $m_h$ increases. The negative contribution from $M_T^2$ to $m^2$ must be compensated by the other positive contributions. Typically, this
implies a large value of the triplet mass,
$M_\phi^2=(\lambda_a+\lambda_b)f^2$, which implies a large value of
$(\lambda_a+\lambda_b)$, but keeping
$1/\lambda=1/\lambda_a+1/\lambda_b$ fixed for a given $m_h$. There are
two ways of getting this:
\bea
\label{ab}
{\mathrm  a)}&&\;\; \lambda\simeq \lambda_b\ll \lambda_a\simeq
M_\phi^2/f^2\ ,\nonumber\\ {\mathrm b)}&&\;\; \lambda\simeq
\lambda_a\ll \lambda_b\simeq  M_\phi^2/f^2\ .  \eea
Notice that the one-loop $m^2$ is a symmetric function of $\lambda_a$
and $\lambda_b$, so cases ${\mathrm a)}$ and ${\mathrm b)}$ are simply
related by $\lambda_a\leftrightarrow\lambda_b$.  In both cases the
triplet and Higgs masses are the same although
the fine-tuning may be different since the dependence of
$\lambda_{a,b}$ on $p_i$ is not the same.

\begin{figure}
\begin{center}
\psfig{figure=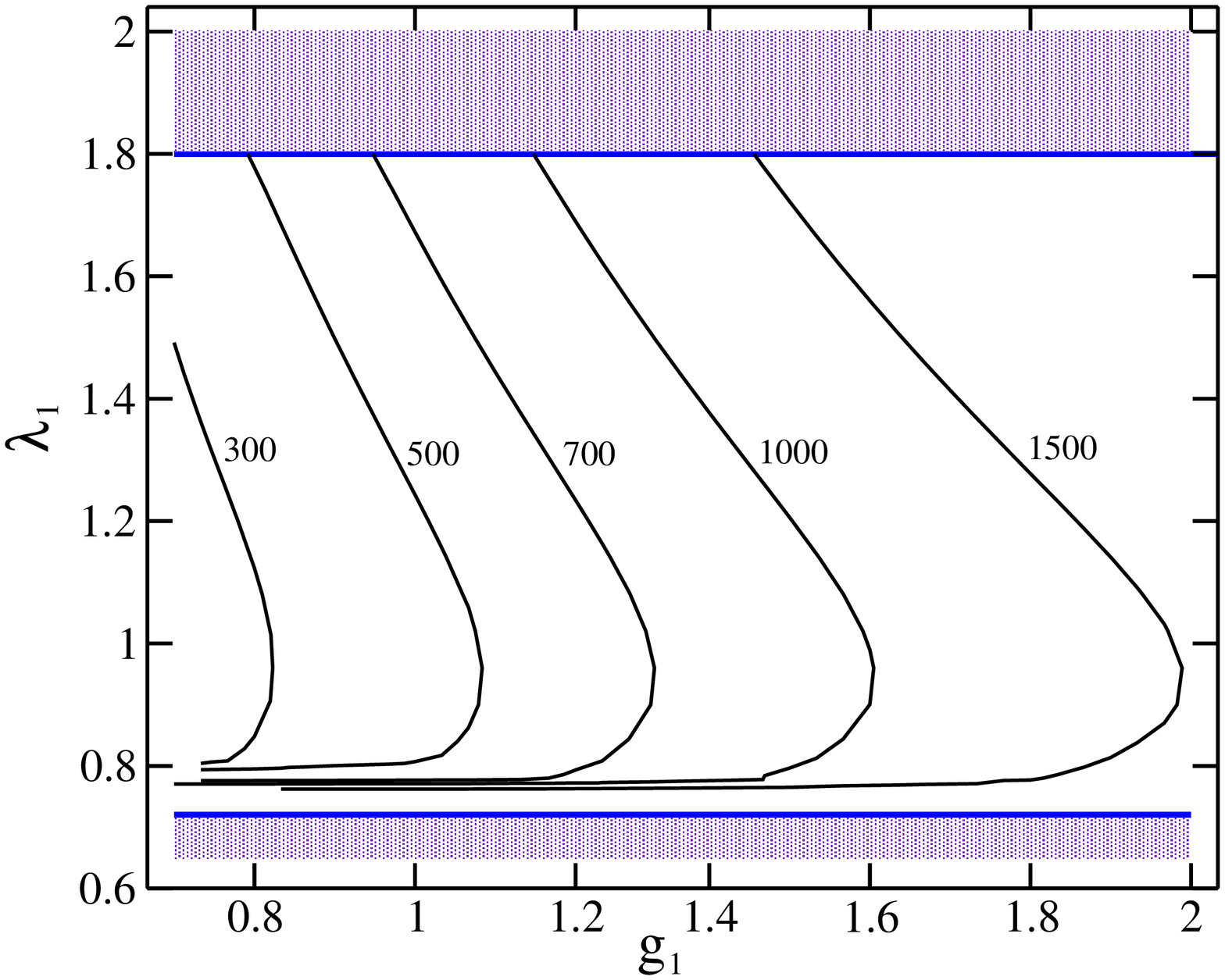,height=3.1in,angle=-90} \hspace{-.65cm}
\psfig{figure=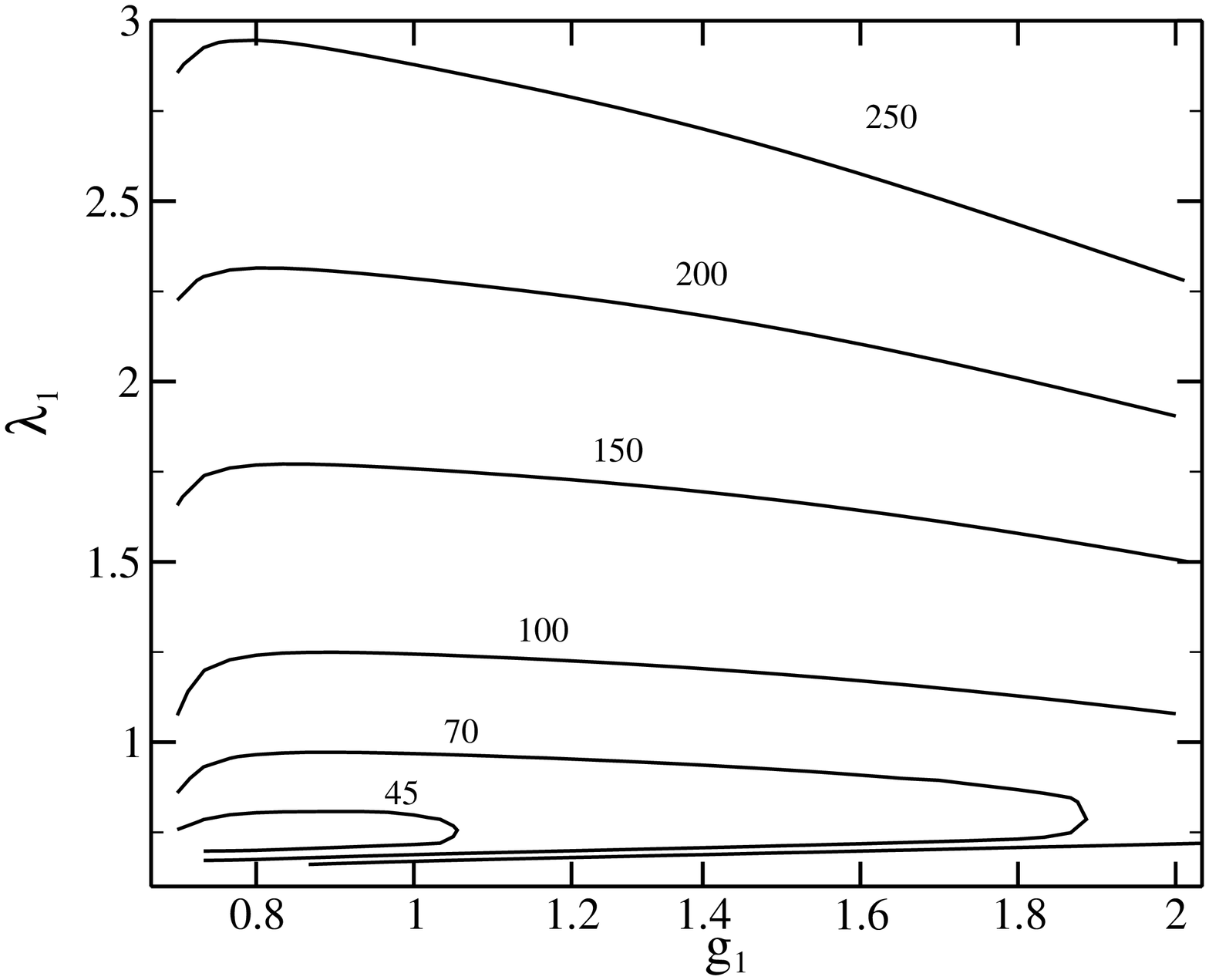,height=3.1in,angle=-90}
\caption{Fine-tuning contours for the Littlest Higgs model, case a), 
for two diferent values of the Higgs mass: $m_h=115$ GeV (left) and 
$m_h=250$ GeV (right).
\label{fig:littlest-a}}
\end{center}
\end{figure}

For case a), the value of $\Delta$ is shown in the contour  plots of
fig.~\ref{fig:littlest-a} which correspond to two different  values of
the  Higgs mass. The results are presented in the plane {$g_1$, $\lambda_1$} \footnote{The value of $g_1'$ has been fixed at $g_1'^2=g_2'^2
= g'^2/2$, which nearly minimizes the fine-tuning. Note also that
$g_1\geq g$. Also $f$ is fixed at 1 TeV. For other values of $f$ the dependence is $\Delta \propto f^2$}. These plots illustrate the large size of $\Delta$, always above ${\cal O}(10)$. This is because besides the heavy top
contribution to $m^2$, there are
other contributions that depend in various ways on the different
independent parameters. We can see from the condition a) in (\ref{ab}),
$\lambda\simeq c(g_1^2+g_1'^2)=\lambda_b\ll
\lambda_a=c(g_2^2+g_2'^2)-c'\lambda_1^2$, that in this
case $c'$ is large (and  negative), while $c$ is small. Then, there is an implicit tuning between $c_0$ and  $c_1$ to get the small value of $c$.  Thus, it makes more sense
to use $\{c_0,\ c_0'\}$, rather than $\{c,\  c'\}$ as the independent
unknown parameters.

For case b) things are much worse. The reason is the following. In case ${\mathrm b)}$, both
$c$ and $c'$ are sizeable so there is no implicit  tuning between
$c_0$ ($c_0'$) and $c_1$ ($c_1'$) but this implies a cancellation to
get $\lambda_a\simeq \lambda$, which requires a delicate tuning. This
``hidden fine-tuning" is responsible for the unexpectedly large values
of $\Delta$. In other words, small changes in the independent
parameters of the model produce large changes in the value of
$\lambda$, and thus in the value of $v^2$.

\subsection{The Simplest Little Higgs Model}

This model \cite{Sch} is based on a global
$[SU(3)\times U(1)]^2/[SU(2)\times U(1)]^2$. The initial gauged
subgroup is   $[SU(3)\times U(1)_X]$ that gets broken to the
electroweak subgroup. This symmetry breaking is now triggered by the VEVs $f_1$ and 
$f_2$ of two $SU(3)$ triplets, $\Phi_1$ and $\Phi_2$. For later use we 
define $f^2\equiv f_1^2+f_2^2$ that measures the total amount of breaking. This spontaneous breaking 
produces 10 Goldstone bosons, 5 of 
which are eaten by the Higgs mechanism to make massive a complex $SU(2)$ 
doublet of extra $W'$s, $(W'^{\pm}, W'^0)$, and an extra $Z'$. The 
remaining 5 degrees of freedom are: $H$ [an $SU(2)$ doublet to be 
identified with the SM Higgs] and $\eta$ (a singlet). The initial tree-level Lagrangian has a structure similar to the one of the Littlest model. In this model the cancellation of $h^2$ terms in ${\rm Str} M^2$  holds to all orders in $h$. Therefore, and in contrast with the Littlest, 
one-loop quadratically divergent corrections from gauge or fermion loops 
do not induce scalar operators to be added to the Lagrangian. Then, no 
Higgs quartic coupling is present at this level.

Less divergent one-loop corrections induce both a mass term and a 
quartic coupling for the Higgs. Using again the $\overline{\rm MS}$ 
scheme and setting the renormalization 
scale $Q=\Lambda$, the one-loop potential on powers of $h$ is \cite{Sch}:

\be
\label{Vrad}
V(h)={1\over 2} (\delta m^2 + \mu_0^2 )h^2 +{1\over 4}\left[\delta_1 
\lambda(h)-{\delta m^2\over 3} {f^2\over f_1^2 f_2^2}-{1\over 24} {\mu_0^2 f^2\over f_1^2 f_2^2}\right]h^4 
+...
\ee
with
\bea
\delta m^2 & = & {3\over 32 \pi^2} \left[g^2 
M_{W'}^2\left(\log{\Lambda^2\over M_{W'}^2}+{1\over 3}\right)
+{1\over 2}(g^2+g'^2)M_{Z'}^2\left(\log{\Lambda^2\over M_{Z'}^2}+{1\over 
3}\right)\right]
\nonumber\\
&-&{3\over 8 \pi^2}\lambda_t^2
M_{T}^2\left(\log{\Lambda^2\over M_{T}^2}+1\right)
+...
\ ,
\label{dm2}
\eea
and
\bea
\delta_1\lambda(h)&=& - {3\over 128 \pi^2} \left[g^4
\left(\log{M_{W'}^2\over m_{W}^2(h)}-{1\over 2}\right)
+{1\over 2}(g^2+g'^2)^2\left(\log{M_{Z'}^2\over m_{Z}^2(h)}-{1\over
2}\right)\right]
\nonumber\\
&+&{3\over 16 \pi^2}\lambda_t^4
\left(\log{M_{T}^2\over m_{t}^2(h)}-{1\over 2}\right)+...
\ .
\label{deltal}
\eea

where a mass term, $\mu_0^2$, is added to the tree-level potential in order to have a positive mass for the Higgs choosing $\mu^2_0>0$. The input parameters are now $\lambda_1$, $\lambda_2$, $\mu^2$, $f_1$ and $f_2$. Without loss of generality we can choose $f_1\leq f_2$, in which case the 
UV cut-off is $\Lambda=4\pi f_1$.
\begin{figure}
\begin{center}
\psfig{figure=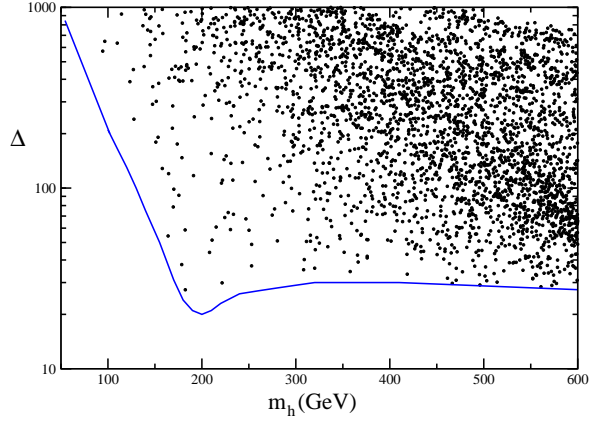,height=3.5in,angle=-90}
\caption{Scatter-plot of the fine-tuning in the Simplest Little Higgs model as a function of the Higgs mass.
\label{fig:scatter}}
\end{center}
\end{figure}
The scatter-plot of fig.~{\ref{fig:scatter}} shows 
the value of $\Delta$ vs. $m_h$ for random values of the parameters compatible with $v=246$ GeV. We have set $f_1=f_2=1$ TeV and chosen at random $\lambda_0\in [-2,2]$, $\lambda_1\in [\lambda_t/\sqrt{2},15]$ and $c_0\in [-10,10]$. The solid line gives minimum values of $\Delta$ and it is clear from the plot only a very small area of the parameter space is closed to this lower bound. Then, we can conclude that the fine-tuning in this model is similar to that of the Littlest: it is always important
and usually comparable (or higher) to that
 of the Little Hierarchy problem [$\Delta \simgt {\cal O}(100)$].

\section{Conclusions}

We have analyzed the fine-tuning associated to the EW breaking process in two Little Higgs (LH) models: the Littlest Higgs \cite{Littlest} and the Simplest Little Higgs \cite{Sch}.

The first conclusion is that these models have a higher fine-tuning than suggested by rough estimates. This is due to implicit tunings between parameters that show up in a more systematic analysis. These implicit tunings are also because of the great amount of superstructure of these models.

The two LH scenarios analyzed present a fine-tuning bigger than 10 \% in most of their parameter space, and the same happens in other models analyzed in ref.~\cite{FT2}. Actually, the fine-tuning is comparable or higher than the one associated to the Little Hierarchy problem of the SM. This unexpected high fine-tuning is mostly because of two reasons. First, the LH models have operators in their lagrangian with the same structure as the operators generated through the quadratic radiative corrections to the
potential. These operators have two contributions: the radiative one (computable)
and the 'tree-level' one (arising from physics beyond the cut-off and
unknown). The required value of the coefficient in front
of a given operator is often much smaller than the calculable contribution,
which implies a  tuning between the tree-level and the one-loop pieces.
Second, the value of the Higgs quartic coupling, $\lambda$, receives
several contributions which have a non-trivial dependence on the various
parameters of the model. Therefore, keeping $\lambda$ in a phenomenologically acceptable region needs an extra fine-tuning.

\section*{Acknowledgments}
I would like to thank A. Casas and J.R. Espinosa for their invaluable help. I would also like to thank Sacha Davidson for her invitation to Moriond.

\end{document}